\documentclass[prd,showpacs,twocolumn,superscriptaddress,floatfix,nofootinbib,10pt]{revtex4-2}
\usepackage{setspace}
\usepackage[utf8]{inputenc}
\usepackage{float}
\usepackage{amsmath,amssymb,amsfonts,bm}
\usepackage{graphicx}
\usepackage{multirow}
\usepackage[usenames,dvipsnames]{color}
\allowdisplaybreaks
\usepackage[
colorlinks=true,
linkcolor=blue,
breaklinks=true,
urlcolor=blue,
citecolor=blue]{hyperref}

\usepackage{orcidlink}
\usepackage{subfigure}
\usepackage{soul}
\usepackage{xcolor}
\usepackage{physics}
\usepackage{booktabs}
\usepackage{ulem}

\definecolor{green}{rgb}{0,0.6,0}
\newcommand{\mev}{\textrm{ MeV}}

\newcommand{\GXNU}{\affiliation{Department of Physics, Guangxi Normal University, Guilin 541004, China}}

\newcommand{\GXZD}{\affiliation{Guangxi Key Laboratory of Nuclear Physics and Technology, Guangxi Normal University, Guilin 541004, China}}

\newcommand{\CSU}{\affiliation{School of Physics, Central South University, Changsha 410083, China}}

\newcommand{\IFIC}{\affiliation{Departamento de F\'{\i}sica Te\'orica and IFIC, Centro Mixto Universidad de
Valencia-CSIC Institutos de Investigaci\'on de Paterna, Apartado 22085,
46071 Valencia, Spain}}

\begin{document}
\title{$a_0(980)$ production, triangle singularity, and non-$\phi$ background in the $J/\psi \to \phi \eta \pi^0$ reaction}

\begin{abstract}
We study the $J/\psi\to\phi\eta\pi^{0}$ reaction measured recently by the BESIII Collaboration with high precision, paying attention to three important aspects: 1) the production of the $a_0(980)$ in the $\pi^0\eta$ mass distributions, with the typical narrow width observed in isospin violating processes;
2) the origin of two peaks in the $\phi\pi^{0}$ mass distributions that were branded as ``non-$\phi$" contribution in the experimental analysis;
3) the existence of two triangle mechanisms developing a triangle singularity at the same energy where the ``non-$\phi$" contribution peaks in the experiment.
However, we also show that the strength of these peaks is very small relative to the observed ones, a feature that is tied to the experimental technique used to identify the $\phi$ looking at $K^+K^-$ in a narrow window of invariant mass around the $\phi$ mass.
We suggest that these triangle singularities could be observed with other methods to identify the $\phi$.
\end{abstract} 

\author{Hai-Peng Li\orcidlink{0009-0008-2985-3011}}
\GXNU%

\author{Wei-Hong Liang\orcidlink{0000-0001-5847-2498}}%
\email{liangwh@gxnu.edu.cn}
\GXNU%
\GXZD%

\author{Chu-Wen Xiao\orcidlink{0000-0001-5303-8350}}%
\GXNU%
\GXZD%
\CSU%

\author{Eulogio Oset\orcidlink{0000-0002-4462-7919}}%
\email{Oset@ific.uv.es}
\GXNU%
\IFIC%

\maketitle

\section{Introduction}\label{sec:Intr}
The issue of $f_0(980)$-$a_0(980)$ mixing has attracted much attention \cite{Achasov:1979xc,Krehl:1996rk,Kudryavtsev:2000txg,Bayar:2017pzq,Kerbikov:2000pu,Close:2000ah,Grishina:2001zj,Close:2001ay,Kudryavtsev:2002uu,Achasov:2002hg,Achasov:2003se,Achasov:2004ur,Wu:2007jh,Hanhart:2007bd,Wu:2008hx,Aceti:2012dj,Roca:2012cv,Tarasov:2013yma,Sekihara:2014qxa,Wang:2016wpc,Achasov:2017edm,Achasov:2017zhu,Sakai:2017iqs,Bayar:2017pzq,Aliev:2018bln,Liang:2017ijf,Wang:2019tqt,Jing:2019cbw,BESIII:2023zwx}.
The isospin breaking mechanism has been earlier associated with loop effects involving kaons,
since both $f_0(980)$ and $a_0(980)$ couple strongly to $K\bar K$, and the mass difference between the charged and neutral kaons induce an isospin violation effect \cite{Achasov:1979xc,Achasov:2003se}.
This picture has become more apparent with the advent of the chiral unitary approach, where the light scalar mesons are dynamically generated from the pseudoscalar-pseudoscalar meson interaction \cite{Oller:1997ti,Kaiser:1998fi,Markushin:2000fa,Nieves:1998hp}.
In particular, both the $f_0(980)$ and $a_0(980)$ couple strongly to the $K\bar K$ component in isospin $I=0$ and $I=1$ respectively.

The $J/\psi \to \phi\eta\pi^0$ reaction was proposed to see the $f_0(980)$-$a_0(980)$ mixing, suggesting a mechanism based on a triangle mechanism with $J/\psi \to K^* \bar K$, followed by $K^* \to \phi K$ and posterior fusion of $K\bar K$ to give $\pi^0 \eta$ through the $a_0(980)$ resonance \cite{Wu:2007jh}.
The difference of masses between the charged and neutral $K\bar K$ pairs in the loop is what makes the isospin forbidden reaction to occur.
The same mechanism with $K\bar K \to \pi \pi$ produces the $f_0(980)$, which is observed in the $J/\psi \to \phi \pi^+\pi^-$ reaction in Ref.~\cite{BES:2004twe}.
The explicit role of the mass difference in this reaction is explained in Ref.~\cite{Hanhart:2007bd}.
The measurements of the $J/\psi\to\phi\eta\pi^{0}$ reaction were performed in Ref.~\cite{BESIII:2010dhc}, where a clear signal was seen in the $\pi^0 \eta$ mass distribution in the region of the $a_0(980)$.
A quantitative evaluation of the decay rates and line shapes was performed in Ref.~\cite{Roca:2012cv}, showing consistency of the experimental data with predictions based on the chiral unitary approach.
Once again, the mass difference between charged and neutral $K\bar K$ pairs is the key ingredient to explain the data.
Further experimental steps were given in Ref.~\cite{BESIII:2018ozj}, where a clearer $a_0(980)$ signal was observed.

More work in this direction was made in Ref.~\cite{Jing:2019cbw}, where it was shown that in the $J/\psi \to \phi \pi^0 \eta$ reaction a triangle singularity (TS) should be seen in the $\phi\pi^0$ invariant $M_{\phi \pi^0}$ around $1385 \mev$, associated with the triangle loop mechanism of Ref.~\cite{Wu:2007jh} discussed above.
Triangle singularities (TSs) appear when in a triangle loop mechanism all intermediate particles can be put on shell and are collinear \cite{Karplus:1958zz,Landau:1959fi} and, in addition, the process can occur at the classical level (Coleman-Norton theorem \cite{Coleman:1965xm}).
A pedagogical and practical approach to TS is given in Ref.~\cite{Bayar:2016ftu} and a recent review on this issue can be seen in Ref.~\cite{Guo:2019twa}.
Actually, in isospin violating processes proceeding through triangle mechanisms, the amount of isospin violation is notably enhanced by the presence of a TS \cite{Sakai:2017iqs,Liang:2017ijf}.
In the present case, since the TS comes from placing intermediate particles on shell, in particular, the kaons of the loop, the effects of the mass difference between charged and neutral kaons are magnified with respect to the cases where the particles are off shell, which are more insensitive to the values of the mass.

The idea developed in Ref.~\cite{Jing:2019cbw} stimulated further experimental work, and the BESIII Collaboration recently presented results on the $J/\psi \to \phi \pi^0 \eta$ reaction with high precision in Ref.~\cite{BESIII:2023zwx}.
In this experiment, the $\phi$ was identified by looking at the invariant mass distribution of $K^+ K^-$ pairs, since $\phi \to K^+ K^-$ is the strongest $\phi$ decay mode.
Interestingly, a huge peak was observed in the $\phi \pi^0$ mass distribution, exactly at the position where a TS was predicted in Ref.~\cite{Jing:2019cbw}.
However, the analysis of Ref.~\cite{BESIII:2023zwx} identified this peak as a non-$\phi$ background contribution, rather than a TS.
The idea is that apart from the mechanism leading to $J/\psi \to \pi^0 \eta \phi, \phi \to K^+K^-$, which violates isospin, there must be other $J/\psi \to \pi^0 \eta K^+ K^-$ processes, with $K^+ K^-$ produced in $I=1$, and, hence, without isospin violation, which contribute to the reaction when a cut $(m_{\phi}-10 \mev) < M(K^+K^-)< (m_{\phi}+10 \mev)$ is performed to identify the $\phi$, as done in Ref.~\cite{BESIII:2023zwx}.
One might ask whether the appearance of the peak of the non-$\phi$ contribution at the same position where the TS is predicted is an accident or a necessary consequence of the dynamics of this non-$\phi$ contribution.
This is an issue that we want to clarify in the present paper.
In addition, we shall also discuss the TS mechanism and link it to the $a_0(980)$ production.
From this perspective, we shall be able to suggest the type of measurement that should be done to eliminate the non-$\phi$ contribution and isolate the TS, which can further clarify the mechanism of isospin violation to produce the $a_0(980)$ and the role that the TS plays in it.

\section{formalism}
\subsection{$J/\psi$ decay to a vector and two pseudoscalars}\label{subsec:JpsiDecay}
We begin by considering $J/\psi$ as a $c\bar c$ state which corresponds to a scalar in SU(3) ($u, d, s$ quarks).
It is easy to see the combination of two pseudoscalars ($PP$) and one vector ($V$) that can come with this symmetry.
For this, we write the SU(3) $q\bar q$ matrices in terms of physical mesons as
\begin{equation}\label{eq:Pmatrix}
   P=
    \left(
    \begin{array}{ccc}
    \frac{1}{\sqrt{2}}\pi^0 + \frac{1}{\sqrt{3}}\eta & \pi^+ & K^+ \\[2mm]
    \pi^- & -\frac{1}{\sqrt{2}}\pi^0 + \frac{1}{\sqrt{3}}\eta & K^0 \\[2mm]
    K^- & \bar{K}^0 & ~-\frac{1}{\sqrt{3}}\eta \\
    \end{array}
    \right),
\end{equation}
\begin{equation}\label{eq:Vmatrix}
    V =
    \left(
    \begin{array}{cccc}
    \frac{1}{\sqrt{2}}\rho^0 + \frac{1}{\sqrt{2}}\omega  & \rho^+ & K^{*+} \\[2mm]
    \rho^- & -\frac{1}{\sqrt{2}}\rho^0 + \frac{1}{\sqrt{2}}\omega  & ~K^{*0}\\[2mm]
    K^{*-} & \bar{K}^{*0} & \phi\\
    \end{array}
    \right),
\end{equation}
where in $P$ we have taken the standard $\eta-\eta'$ mixing of Ref.~\cite{Bramon:1992kr} and ignored the $\eta'$ that plays no role in the present process.
The SU(3) scalars involving $PPV$ can be written in terms of these matrices as
\begin{equation}
	\langle V PP\rangle , ~~\langle V \rangle  \langle PP\rangle , ~~\langle VP\rangle \langle P\rangle, ~~\langle V\rangle \, \langle P\rangle  \langle P\rangle,
\end{equation}
where $\langle \, \rangle$ indicates the trace of these SU(3) matrices.
However, arguments of heavy quark symmetry indicate that any extra trace induces a reduction factor of order $\frac{1}{N_c}$ \cite{Manohar:1998xv}, where $N_c$ is the number of colors.
This indicates that the $\langle V PP\rangle$, $\langle V \rangle \langle PP\rangle$, and $\langle V P\rangle \, \langle P\rangle$ structures are the dominant ones and one can neglect the $\langle V\rangle \, \langle P\rangle  \langle P\rangle$ structure.
In Ref.~\cite{Jing:2019cbw} the $\langle VPP \rangle$ structure is used.
In the study of $J/\psi \to \eta h_1, \pi^0 b_1$ decays in Ref.~\cite{Liang:2019vhf}, the $\langle V PP\rangle$ and $\langle V \rangle  \langle PP\rangle$ structures were used.
Also in the study of $\chi_{c1} \to \eta \pi^+ \pi^-$ in Refs.~\cite{Liang:2016hmr,Debastiani:2016ayp}, the term with three traces was found to lead to unacceptable shapes.

Following Ref.~\cite{Liang:2019vhf}, we observe that the $\langle V PP\rangle$ structure does not contain any $\phi \pi^0 \eta$ term since it violates isospin.
Instead, it contains the useful term
\begin{equation}\label{eq:Hadron1}
	H= \phi \;(K^- K^+ +\bar K^0 K^0),
\end{equation}
which, with our isospin phase convention of $(K^+,K^0)$ and $(\bar K^0, -K^-)$, corresponds to a $\phi (K\bar K, I=0)$ state, which preserves isospin.
The spin structure of the $J/\psi$ and $\phi$ is accounted for by means of a coupling~\cite{Jing:2019cbw}
\begin{equation}\label{eq:t1}
	-it_{J/\psi \to \phi K\bar K}=A_1\, \epsilon^\mu (J/\psi) \, \epsilon_\mu (\phi),
\end{equation}
with $A_1$ a constant.

\begin{figure}[t]
\centering
\includegraphics[width=0.33\textwidth]{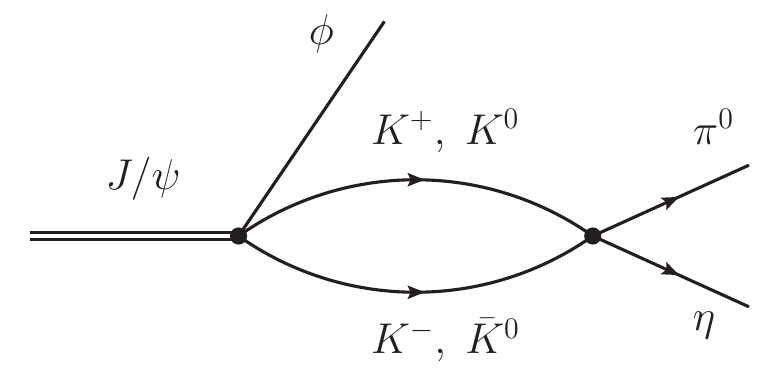}
\vspace{-0.3cm}
\caption{Loop mechanism involving FSI of $K\bar K$ for $J/\psi \to \phi \pi^0 \eta$.}
\label{Fig:Fig1}
\end{figure}

It is now easy to see how the $J/\psi \to \phi \pi^0 \eta$ can be obtained.
Indeed, if we allow for final state interaction (FSI) of the $K\bar K$ in Eq.~\eqref{eq:Hadron1} we have the mechanism depicted in Fig.~\ref{Fig:Fig1}.
The amplitude for the decay process depicted in Fig.~\ref{Fig:Fig1} is given by
\begin{align}\label{eq:ta0}
	t_{a_0}
	=&A_1\, \epsilon^\mu (J/\psi) \, \epsilon_\mu (\phi) \;\Big\{G_{K^+K^-}(M_{\rm inv})\; t_{K^+K^-,\pi^0\eta}(M_{\rm inv})
	\nonumber \\
		 &+G_{K^0\bar{K}^0} (M_{\rm inv})\; t_{K^0\bar{K}^0,\pi^0\eta}(M_{\rm inv})
		 \Big\},
\end{align}
where $M_{\rm inv}\equiv M_{\rm inv} (\pi^0 \eta)$, $G_{K\bar K}$ are the loop functions of $K\bar K$ in Fig.~\ref{Fig:Fig1}, and $t_{i,j}$ are the scattering matrices of two pseudoscalars evaluated within the chiral unitary approach \cite{Oller:1997ti,Kaiser:1998fi, Markushin:2000fa,Nieves:1998hp}.
We take explicitly into account the $\eta-\eta'$ mixing of Ref.~\cite{Bramon:1992kr} in the transition potentials and take the scattering matrices from Ref.~\cite{Lin:2021isc}.
The loop functions $G_{K\bar K}$ are regularized with the cutoff method in Ref.~\cite{Lin:2021isc} with a cutoff $q_{\rm max}=630 \mev$.

If the masses of $K^+, K^-$ were the same as those of $K^0, \bar K^0$, then the two terms in Eq.~\eqref{eq:ta0} cancel, the $|K\bar K, I=0 \rangle $ state then cannot lead to $\pi^0 \eta$, and isospin symmetry is preserved.
However, if one takes the charged and neutral physical kaon masses, the loops do not cancel, leading to a violation of isospin.
Actually, since the $t_{i,j}$ matrices are constructed through $T=[1-VG]^{-1}\, V$ and involve the $G$ loops, there is also an isospin violation in the $t_{i,j}$ matrices, and the final isospin violation in $J/\psi \to \phi \pi^0 \eta$ relies upon the two sources.

In Ref.~\cite{Liang:2019vhf} the $\langle V \rangle  \langle PP\rangle$ structure is also evaluated and one finds a structure
\begin{equation} \label{eq:Hadron2}
	H'=2 \, \phi \;(K^+K^- +K^0 \bar K^0),
\end{equation}
which would have a weight $2A_1 \beta$, with $\beta$ on the order of $N_c^{-1}$ according to Ref.~\cite{Manohar:1998xv}.
The structure is the same as that of Eq.~\eqref{eq:Hadron1}; hence we automatically can take it into account by substituting in Eq.~\eqref{eq:t1}
\begin{equation}
	A_1 \to A_1\, (1+2\beta).
\end{equation}
We take $\beta=0.0927$ from Ref.~\cite{Roca:2004uc}, from the study of $J/\psi\to\phi\pi\pi$ and $J/\psi\to\omega\pi\pi$.

\subsection{Triangle singularity}
\begin{figure}[b]
\centering
\includegraphics[width=0.48\textwidth]{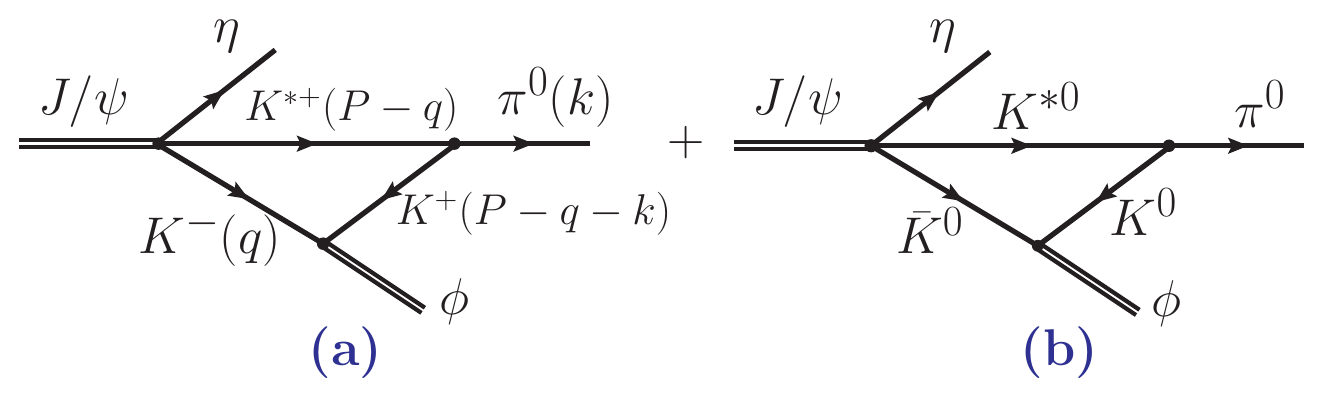}\\[1mm]
\includegraphics[width=0.48\textwidth]{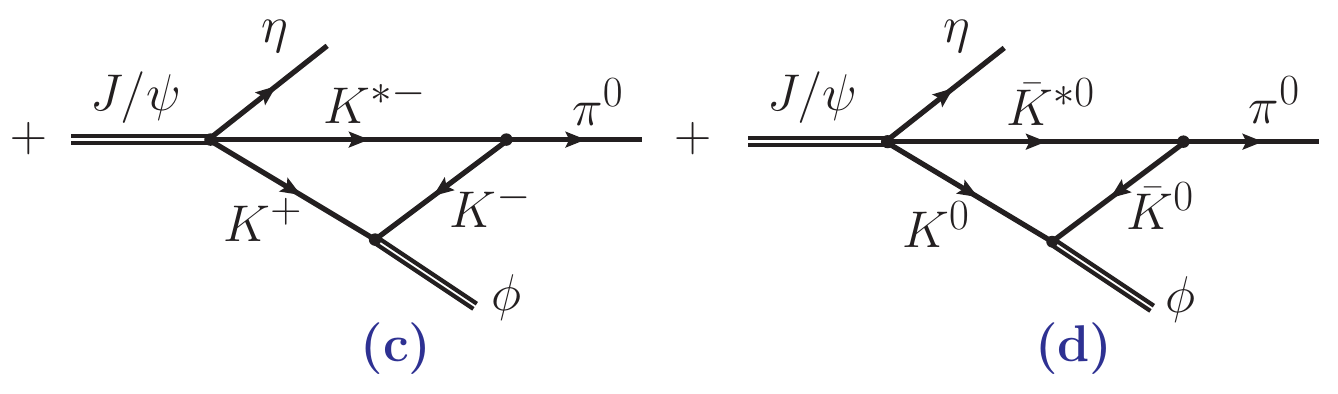}
\vspace{-0.4cm}
\caption{Triangle diagrams for $J/\psi \to \phi \pi^0 \eta$ decay. (a), (b), (c), and (d) correspond to different charges in the intermediate states. The momenta of the particles are written in parentheses. $P$ is the sum of the $\pi^0$ and $\phi$ momenta. $K^* \equiv K^*(890)$.}
\label{Fig:Fig2}
\end{figure}

We now look at the triangle mechanism proposed in Ref.~\cite{Jing:2019cbw} that develops a TS.
The mechanism is depicted in Fig.~\ref{Fig:Fig2}.

The $K^* \to \pi K$ and $\phi \to K\bar K$ vertices are evaluated from the standard $VPP$ Lagrangian
\begin{equation}
	\mathcal{L}=-ig\, \langle [ P, \,\partial_\mu P] V^\mu\rangle,
\end{equation}
with $g=\dfrac{M_V}{2f}, M_V=800 \mev$, and $f=93 \mev$.
The $J/\psi \to K^* \eta \bar K$ vertex is now obtained from the $\langle VP\rangle \langle P\rangle$ structure and we find
\begin{equation}
	H^{''}=\dfrac{\eta}{\sqrt{3}} \, (K^{*+} K^- +K^{*0}\bar K^0 +K^{*-} K^0 + \bar K^{*0} K^0),
\end{equation}
to which we associate a weight $A_1 \alpha$, where, according to Ref.~\cite{Manohar:1998xv}, $\alpha$ would also be of the order of $N_c^{-1}$.
Yet, it is interesting to note that in Ref.~\cite{Jing:2019cbw} this decay was obtained from the $\langle VPP\rangle$ structure.
We find that in our approach the $\langle VPP\rangle$ structure does not lead to this decay and the reason is that we use the $\eta-\eta'$ mixing of Ref.~\cite{Bramon:1992kr}, while in Ref.~\cite{Jing:2019cbw} the $\eta$ is assumed to be a member of the SU(3) octet.
This means that a small deviation of the mixing of Ref.~\cite{Bramon:1992kr} will make the $\langle VPP\rangle$ structure contribute and, hence, we should not expect the $N_{c}^{-1}$ reduction.
We also find that the $\langle V\rangle\langle PP\rangle$ structure does not lead to the $J/\psi\to \bar{K}^{*}K\eta$~\cite{Liang:2019vhf}.
Thus, the $\langle VP\rangle\langle P\rangle$ structure is the only one that contributes to $J/\psi\to \bar{K}^{*}K\eta$ in our approach.

The calculation of the mechanism of Fig.~\ref{Fig:Fig2} is notably simplified if we consider only the spatial components of the $\phi, K^*$ polarization vectors, which is well justified in this case, as one can see from the arguments of Refs.~\cite{Sakai:2017hpg,Dias:2025izv}.
We evaluate the loops in the frame of reference where the $\pi^0 \phi$ system is at rest.
Thus, taking $P=p_{\pi^0}+p_\phi$, $\vec P=0$.
All the $VPP$ vertices have now the structure
\begin{equation}\label{eq:tVPP}
	-it_{VPP}^{(j)}= -i C^{(j)}\, g \,\epsilon^i (V) \, (2k+q)^i,
\end{equation}
with the coefficients $C^{(j)}$ given in Table \ref{Tab:tab1}.
\begin{table*}[!t]
\centering
\caption{$C^{(j)}$ coefficients of Eq.~\eqref{eq:tVPP}. In parentheses we denote the diagram in Fig.~\ref{Fig:Fig2}.}
\label{Tab:tab1}
\setlength{\tabcolsep}{10pt}
\begin{tabular}{c|cccc}
\hline\hline
Vertex & $K^{*+}\to K^+ \pi^0$ & $K^{*-}\to K^- \pi^0$ & $K^{*0} \to K^0 \pi^0$ &$\bar K^{*0} \to \bar K^0 \pi^0$\\
\hline
$C^{(j)}$ & $\dfrac{1}{\sqrt{2}}$ & $-\dfrac{1}{\sqrt{2}}$ & $-\dfrac{1}{\sqrt{2}}$ & $\dfrac{1}{\sqrt{2}}$ \\[3mm]
\hline\hline
Vertex & $K^- K^- \to \phi$ (a) & $K^- K^- \to \phi$ (c) & $\bar K^0 K^0 \to \phi$ (b) & $K^0 \bar K^0 \to \phi$ (d)\\
\hline
$C^{(j)}$ & $1$ & $-1$ & $1$ & $-1$\\
\hline\hline
\end{tabular}
\end{table*}

One can now see, by looking at Eq.~\eqref{eq:tVPP} and Table \ref{Tab:tab1}, that Fig.~\ref{Fig:Fig2}(a) and Fig.~\ref{Fig:Fig2}(c) give equal contribution and Fig.~\ref{Fig:Fig2}(b) and Fig.~\ref{Fig:Fig2}(d) also provide the same contribution.
The two blocks have, however, opposite sign and if the masses of the charged and neutral kaons are equal, and also those for the $K^*$, then the four diagrams cancel, as should be the case for an isospin conserving process.
However, once again, if we consider the physical masses of the kaons the cancellation does not occur and we have an isospin violating process.

There is another simplifying method to write the amplitude for the triangle diagram.
Since the TS develops when all three intermediate propagators are placed on shell, it is sufficient to take the positive energy part of the propagators.
This means that we write
\begin{equation}
	\dfrac{1}{q^2-m^2}= \dfrac{1}{2\,\omega(\vec q\,)} \left(\dfrac{1}{q^0-\omega(\vec q\,)+i\varepsilon} - \dfrac{1}{q^0+\omega(\vec q\,)-i\varepsilon} \right),
\end{equation}
with $\omega(\vec q\,)=\sqrt{\vec q^{\;2} +m^2}$, and take only the first part when $q^0$ is positive.
With this and taking into account that Fig.~\ref{Fig:Fig2}(a) and Fig.~\ref{Fig:Fig2}(c) are equal and Fig.~\ref{Fig:Fig2}(b) and Fig.~\ref{Fig:Fig2}(d) are also equal and of opposite sign to Fig.~\ref{Fig:Fig2}(a) and Fig.~\ref{Fig:Fig2}(c), we can write
\begin{equation}
	t_{\rm TS}= 2\, t_c -2\, t_n,
\end{equation}
where $t_n$ has the same structure as $t_c$, except the physical masses of the neutral particles are used, while in $t_c$ the physical charged masses are used.
Then we have
\begin{align}\label{eq:tc1}
	-it_c= &A_1\alpha \dfrac{1}{\sqrt{3}}
	\int \dfrac{d^4q}{(2\pi)^4}\, \dfrac{1}{\sqrt{2}} g^2 (-i) \epsilon^m (J/\psi) \epsilon^m(K^*) \nonumber \\[1mm]
		&\cdot \dfrac{1}{2\, \omega_K(\vec q\,)}\, \dfrac{1}{2\, \omega^*(\vec q\,)}\, \dfrac{1}{2\, \omega_K(\vec q+\vec k\,)}\, \epsilon^i (K^*) (2k+q)^i \nonumber\\[1mm]
		&\cdot \epsilon^j(\phi) (2q+k)^j \cdot \dfrac{i}{P^0-q^0-\omega^*(\vec q\,)+i\frac{\Gamma_{K^*}}{2}} \nonumber\\[1mm]
		&\cdot \dfrac{i}{P^0-q^0-k^0-\omega_K(\vec q+\vec k\,)+i\varepsilon} \;\dfrac{i}{q^0-\omega_K(\vec q\,)+i\varepsilon},
\end{align}
where $\omega^*(\vec q\,)=\sqrt{\vec q^{\,2}+ m_{K^*}^2}$ and $\omega_K(\vec q\,)=\sqrt{\vec q^{\,2}+ m_{K}^2}$.
Note also that we have considered explicitly the width of the $K^*$ by substituting $\omega^*(\vec q\,) \to \omega^*(\vec q\,) -i\Gamma_{K^*}/2$.
The $q^0$ integral of Eq.~\eqref{eq:tc1} is performed using Cauchy integration which picks up $q^0=\omega_K(\vec q\,)$ and then we write, taking into account that
\begin{equation}\label{eq:pol}
	\sum_{\rm pol} \epsilon^m(K^*)\, \epsilon^i(K^*)=\delta_{im},
\end{equation}
\begin{align}\label{eq:tc2}
  t_c= &-\dfrac{1}{\sqrt{6}} \,A_1 \,\alpha \,g^2 \;\epsilon^i (J/\psi) \, \epsilon^j(\phi) \nonumber \\
  & \cdot \int \dfrac{d^3q}{(2\pi)^3}\, (2k+q)^i \;(2q+k)^j \; f(\vec q, \vec k\,),
\end{align}
  with
  \begin{align}\label{eq:f}
	f(\vec q, \vec k\,)\equiv &\dfrac{1}{2\, \omega_K(\vec q\,)}\, \dfrac{1}{2\, \omega^*(\vec q\,)}  \dfrac{1}{2\, \omega_K(\vec q+\vec k\,)} \nonumber\\
	&\cdot  \dfrac{1}{P^0-\omega_K(\vec q\,)-\omega^*(\vec q\,)+i\frac{\Gamma_{K^*}}{2}} \nonumber\\
	&\cdot \dfrac{1}{P^0-\omega_K(\vec q\,)-k^0-\omega_K(\vec q+\vec k\,)+i\varepsilon}.
\end{align}
The structure of $t_c$ has become very easy, and from there it is easy to see at which $P^0$ energy the TS appears, as done in Ref.~\cite{Bayar:2016ftu} [see Eq.~(18) of that reference].
We find that a TS develops at $P^0=1385\mev$, as also found in Ref.~\cite{Jing:2019cbw}.
When the $K^*$ width is omitted, the amplitude develops a singularity, which turns into a pronounced finite peak when the $K^*$ width is explicitly considered.
We should also note that in all cases $t_c$ is logarithmically divergent.
However, there is no need to regularize it because we need $t_c-t_n$ and in the difference the divergence disappears, since for large values of $\vec q$ the masses in the integrand do not matter.
Next, by taking into account that
\begin{equation}\label{eq:integral}
\begin{aligned}
    & \int d^3q \; q^i \,f(\vec q, \vec k\,) = a k^i, \\
    & \int d^3q \;q^i \,q^j = a' \delta_{ij} +b' k^i k^j,
\end{aligned}
\end{equation}
we obtain
\begin{equation}\label{eq:tc3}
  t_c= -\dfrac{1}{\sqrt{6}} \,A_1 \,\alpha \,g^2 \;\epsilon^i (J/\psi) \; \epsilon^j(\phi)\, \left[ A\, \delta_{ij} +B\, k^i k^j  \right],
\end{equation}
where
\begin{equation}\label{eq:abab}
\begin{aligned}
  A&=2a', \\
  B&=5a+2b+2b',\\
  a&= \int \dfrac{d^3q}{(2\pi)^3} \left(\dfrac{\vec q \cdot \vec k}{\vec k^2}\right)\;f(\vec q, \vec k\,), \\
  b&= \int \dfrac{d^3q}{(2\pi)^3} \;f(\vec q, \vec k\,) ,\\
  a'&= \dfrac{1}{2\, {\vec k}^{\,2}}\;\int \dfrac{d^3q}{(2\pi)^3} \;f(\vec q, \vec k\,) \; \left[ {\vec q}^{\,2} \,{\vec k}^{\,2} - (\vec q \cdot \vec k)^2 \right],\\
  b'&= \dfrac{1}{2\, {\vec k}^{\,4}}\;\int \dfrac{d^3q}{(2\pi)^3} \;f(\vec q, \vec k\,) \; \left[ 3\, (\vec q \cdot \vec k)^2 -{\vec q}^{\,2} \,{\vec k}^{\,2} \right].
\end{aligned}
\end{equation}

\subsection{Non-$\phi$ contribution in $J/\psi \to \pi^0 \eta K^+ K^-$ decay}
\begin{figure}[b]
\centering
\includegraphics[width=0.5\textwidth]{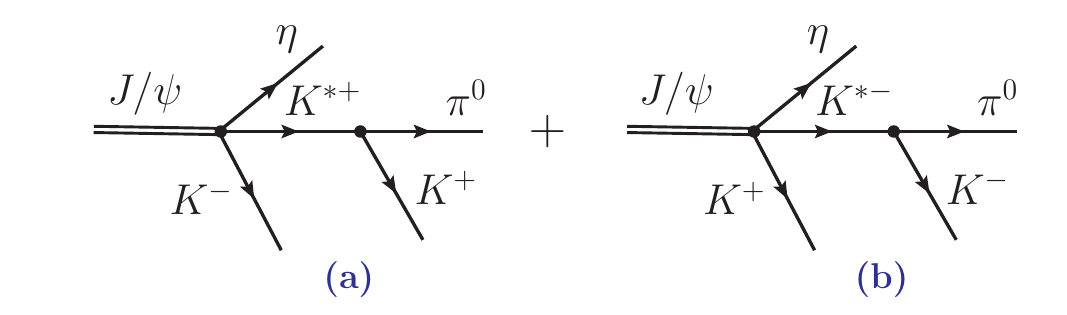}
\vspace{-0.45cm}
\caption{Tree level contribution for $J/\psi \to \pi^0 \eta K^+ K^-$ decay (non-$\phi$ contribution), with $K^* \equiv K^*(890)$. (a) and (b) correspond to different charges in the intermediate states.}
\label{Fig:Fig3}
\end{figure}

Since the $\phi$ is identified in the BESIII experiment through the $\phi$ decay into $K^+K^-$, with the cut
\begin{equation}\label{eq:BEScut}
	M_\phi -10\mev < M_{\rm inv}(K^+K^-) < M_\phi +10\mev,
\end{equation}
we have unavoidably a contribution from the tree level mechanism associated with Fig.~\ref{Fig:Fig2}, which we depict in Fig.~\ref{Fig:Fig3}.

It is clear that the mechanism of Fig.~\ref{Fig:Fig3} will contribute to the mass distribution when we make the cut of Eq.~\eqref{eq:BEScut}.
Note that now the $K^+K^-$ pair is produced in $I=1$, since there is no isospin violation in that process.
Thus, this mechanism does not lead to $\phi$ production but will contribute to the mass spectra of Ref.~\cite{BESIII:2023zwx}, where indeed a large background below the $\phi$ peak in the $K^+K^-$ mass distribution is observed.
We should also note that since the $J/\psi \to \eta K^* \bar K, \eta \bar K^*  K$ vertices are the same as used to evaluate the TS mechanism, the strength of the mechanisms of Figs.~\ref{Fig:Fig2} and \ref{Fig:Fig3} are related.
Yet, since the TS appears from cancellation of diagrams to get the isospin violation mode, which does not occur in the tree level mechanism that does not violate isospin, we can anticipate that the strength of the non-$\phi$ contribution should be much larger than that of the TS.

There is another issue worth mentioning here related to the Schmid theorem \cite{Schmid:1967ojm} (see also Ref.~\cite{Debastiani:2018xoi} for detailed discussion concerning this theorem), where the TSs are reabsorbed into the tree level by multiplying it by the phase of the $K^+K^-$ scattering amplitude in the present case.
However, since the tree level and the TS correspond to different isospin states, this should not apply here.
In any case, the TS contribution is very small here and we would not have the $K^+K^-$ background if one detected the $\phi$ through a different decay mode, in which case the TS contribution would be the same calculated here.

The amplitude for the tree level of Fig.~\ref{Fig:Fig3}(a) is given by
\begin{align}\label{eq:tta}
  -it_{\rm ta}= &-i A_1 \,\dfrac{\alpha}{\sqrt{3}} \,(-i) \,\epsilon^m (J/\psi) \; \epsilon^m(K^*) \; D_1(K^*) \nonumber \\
  &\cdot (-i) \,\frac{g}{\sqrt{2}} \;\epsilon^i(K^*)\; (p_{\pi^0}-p_{K^+})^i \nonumber \\
  =& iA_1 \dfrac{\alpha}{\sqrt{3}}\; D_1(K^*) \,\frac{g}{\sqrt{2}}\; \epsilon^i (J/\psi) \; (p_{\pi^0}-p_{K^+})^i,
\end{align}
where the second equation holds after summing over the $K^*$ polarizations.
Similarly, for Fig.~\ref{Fig:Fig3}(b) we get
\begin{equation}\label{eq:ttb}
	-it_{\rm tb}=iA_1 \dfrac{\alpha}{\sqrt{3}}\; D_2(K^*) \,\frac{g}{\sqrt{2}}\; \epsilon^i (J/\psi) \; (p_{\pi^0}-p_{K^-})^i.
\end{equation}
In Eqs.~\eqref{eq:tta} and \eqref{eq:ttb}, the $D_1(K^*)$ and $D_2(K^*)$ are given by
\begin{align}\label{eq:D1D2}
    & D_1(K^*)=\dfrac{1}{M^2_{\rm inv}(\pi^0 K^+) - m^2_{K^*}+i m_{K^*} \Gamma_{K^*}}, \\
    & D_2(K^*)=\dfrac{1}{M^2_{\rm inv}(\pi^0 K^-) - m^2_{K^*}+i m_{K^*} \Gamma_{K^*}}.
\end{align}

There is still one further step we must do, since the experiment measures $J/\psi \to \pi^0 \eta K^+ K^-$, with four particles in the final state.
The tree level amplitudes already have the four particles explicitly in the final state, but the $a_0$ and TS amplitudes have been calculated with $\phi$ in the final state.
One must complete these amplitudes by putting explicitly the $\phi$ propagator and its decay into $K^+K^-$.
This is immediately done by multiplying the former formula of Eqs.~ \eqref{eq:ta0} and \eqref{eq:tc3} by the factor
\begin{equation}
	i D_{\phi}\; (-i) g\, \epsilon^i (\phi) \; (p_{K^-}-p_{K^+})^i,
\end{equation}
where
\begin{equation}
	D_\phi= \dfrac{1}{M^2_{\rm inv}(K^+ K^-) - m^2_{\phi}+i m_{\phi} \Gamma_{\phi}}.
\end{equation}
Then $t_{a_0}$ gets converted into $\tilde{t}_{a_0}$,
\begin{align}\label{eq:ta0til}
	\tilde{t}_{a_0}
	=&-A_1\, (1+2\beta)\, \epsilon^i (J/\psi) \, g \,D_\phi \;(p_{K^-}-p_{K^+})^i \nonumber \\[1mm]
	&\cdot \Big\{G_{K^+K^-}(M_{\rm inv}(\pi^0\eta))\; t_{K^+K^-,\pi^0\eta}(M_{\rm inv}(\pi^0\eta))
	\nonumber \\[1mm]
		 &+G_{K^0\bar{K}^0} (M_{\rm inv}(\pi^0\eta))\; t_{K^0\bar{K}^0,\pi^0\eta}(M_{\rm inv}(\pi^0\eta))
		 \Big\},~~~~~
\end{align}
and $t_{c}$ gets converted into $\tilde{t}_{c}$,
\begin{equation}\label{eq:tctil}
  \tilde{t}_c= -\dfrac{A_1}{\sqrt{6}} \, \alpha \,g^3 \;\epsilon^i (J/\psi) \; D_\phi \left[ A\, \delta_{ij} +B\, k^i k^j  \right]  (p_{K^-}-p_{K^+})^j.
\end{equation}
Altogether, the final amplitude is given by
\begin{align}\label{eq:ttil}
	\tilde{t}=&2\,\tilde{t}_{c} -2\, \tilde{t}_{n} +\tilde{t}_{a_0} +t_{\rm ta}+t_{\rm tb}\nonumber \\
	=&\epsilon^i (J/\psi) \; V^i,
\end{align}
with
\begin{align}\label{eq:Vi}
	V^i=& -2\, \dfrac{A_1}{\sqrt{6}} \, \alpha \,g^3 \; D_\phi \, (A_c-A_n)\, (p_{K^-}-p_{K^+})^i \nonumber \\[1mm]
	& -2\, \dfrac{A_1}{\sqrt{6}} \, \alpha \,g^3 \; D_\phi \, (B_c-B_n)\, p^i_{\pi^0} \; \vec p_{\pi^0} \cdot  (\vec p_{K^-}-\vec p_{K^+}) \nonumber \\[1mm]
	&-A_1\, (1+2\beta)\, g \,D_\phi \;(p_{K^-}-p_{K^+})^i \nonumber \\
	&\cdot \Big\{G_{K^+K^-}(M_{\rm inv}(\pi^0\eta))\; t_{K^+K^-,\pi^0\eta}(M_{\rm inv}(\pi^0\eta))
	\nonumber \\[1mm]
	&+G_{K^0\bar{K}^0} (M_{\rm inv}(\pi^0\eta))\; t_{K^0\bar{K}^0,\pi^0\eta}(M_{\rm inv}(\pi^0\eta))
		 \Big\}\nonumber \\[1mm]
	&-A_1 \dfrac{\alpha}{\sqrt{3}}\; D_1(K^*) \,\frac{g}{\sqrt{2}} \; (p_{\pi^0}-p_{K^+})^i\nonumber \\[1mm]
	&+A_1 \dfrac{\alpha}{\sqrt{3}}\; D_2(K^*) \,\frac{g}{\sqrt{2}}\; (p_{\pi^0}-p_{K^-})^i,
\end{align}
where $A_c, A_n$ and $B_c, B_n$ are the expressions of $A$ and $B$ from Eq.~\eqref{eq:abab} for the charged and neutral particles of the triangle loops, respectively.
The average over the $J/\psi$ polarizations of $|\tilde{t}|^2$ is done by using
\begin{equation}
	\frac{1}{3}\sum_{\rm pol} \epsilon^i (J/\psi)\, \epsilon^j (J/\psi) = \frac{1}{3}\, \delta_{ij}.
\end{equation}

\subsection{Tree level with other $K^*$ intermediate states}
In the BESIII paper, it is mentioned that ``the components of the non-$\phi$ background are very complicated due to $K^{*(l)}$ contributions'', but no further details are provided.
Here we have stripped out explicitly one of such contributions in Fig.~\ref{Fig:Fig3}, with the intermediate $K^*$ states being the $K^*(890)$.
From the Particle Data Group \cite{ParticleDataGroup:2024cfk}, we find the next two $K^*(1^-)$ resonances, $K^*(1410)$ and $K^*(1680)$.
We shall see that the peak around $2100 \mev$ in the $\phi\pi^0$ mass distribution of Ref.~\cite{BESIII:2023zwx} is naturally produced with the $K^{*}(1410)$ intermediate state replacing the $K^{*}(892)$ in Fig.~\ref{Fig:Fig3}.
The contribution of the $K^{*}(1410)$ is easily done by substituting
\begin{equation}\label{eq:Kstar2}
\begin{aligned}
    & \dfrac{\alpha}{\sqrt{3}} \; D_1(K^*)  \to  \left[\dfrac{\alpha}{\sqrt{3}} \; D_1(K^*) + \dfrac{\gamma}{\sqrt{3}} \; D'_1(K^*) \right], \\[1mm]
    & \dfrac{\alpha}{\sqrt{3}} \; D_2(K^*)  \to  \left[\dfrac{\alpha}{\sqrt{3}} \; D_2(K^*) + \dfrac{\gamma}{\sqrt{3}} \; D'_2(K^*) \right],
\end{aligned}
\end{equation}
in Eqs.~\eqref{eq:tta} and \eqref{eq:ttb}, where
\begin{equation}\label{eq:Dprime}
\begin{aligned}
    & D'_1(K^*)\\
	=&\dfrac{1}{M^2_{\rm inv}(\pi^0 K^+) - m^2_{K^*(1410)}+i m_{K^*(1410)} \Gamma_{K^*(1410)}}, \\[1.5mm]
    & D'_2(K^*) \\
	=&\dfrac{1}{M^2_{\rm inv}(\pi^0 K^-) - m^2_{K^*(1410)}+i m_{K^*(1410)} \Gamma_{K^*(1410)}}.
\end{aligned}
\end{equation}
The parameters $\alpha$ and $\gamma$ will be fitted to the data of BESIII.

After all this is done, the mass distributions are obtained using Monte Carlo integration as indicated in the Appendix.

To make the discussions clearer in the next section, we will rename the coefficients entering our calculation as
\begin{enumerate}
		\item[1)] $A_1\,(1+\beta)=A_{a_0}$, entering the production of the $a_0(980)$ resonance;
		\item[2)] $A_1\,\alpha = A_{tT, 890}$, accounting for tree level ($t$) and triangle mechanism ($T$) with the $K^*(890)$ intermediate state;
		\item[3)] $A_1\,\gamma = A_{t, 1410}$, accounting for tree level with the $K^*(1410)$ intermediate state.
\end{enumerate}

\section{results}
We do not pretend to make a perfect fit to the data but to understand the origin of the different structures shown by the experiment \cite{BESIII:2023zwx}. These are:
\begin{enumerate}
		\item[1)] Two peaks at low invariant masses in the $\phi\eta$ mass distributions. These were explained in the analysis of Ref.~\cite{BESIII:2023zwx} as coming from $\phi(1680)$ and $h_1(1900)$ resonances, and we do not dispute these claims.
		Then, to account for these two peaks we parametrize the $\phi\eta$ amplitude around these peaks in terms of two Breit-Wigner distributions;
		\item[2)] A pronounced peak in the $\phi \pi^0$ mass distribution around $1400 \mev$, that we fit from the tree level amplitude of Fig.~\ref{Fig:Fig3} with $K^*(890)$ intermediate state;
		\item[3)] A smaller and wider peak around $2100 \mev$ in the $\phi\pi^0$ mass distribution, which we obtain from the tree level amplitude of Fig.~\ref{Fig:Fig3}, with $K^*(1410)$ intermediate state;
		\item[4)] The narrow structure around $980 \mev$, corresponding to the $a_0$ excitation, in the $\eta \pi^0$ mass distribution, which comes from the mechanism of Fig.~\ref{Fig:Fig1}.
\end{enumerate}

\begin{figure}[b]
\centering
\includegraphics[width=0.49\textwidth]{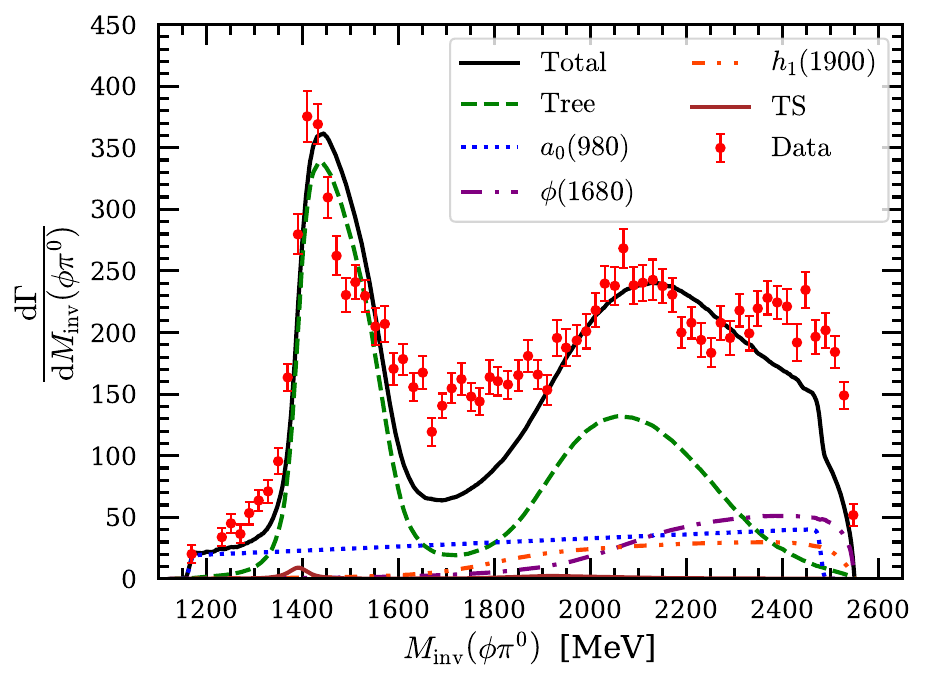}
\vspace{-0.45cm}
\caption{Results for the $\phi\pi^{0}$ invariant mass distribution. The red error bars represent experimental data from Ref.~\cite{BESIII:2023zwx}. The black solid line corresponds to the total result; the green dashed line shows the tree level contribution; the blue dotted line represents the $t_{a_{0}}$ contribution, the brown solid line corresponds to the triangle singularity, and the purple and orange dash-dotted line denote the contributions from the $\phi(1680)$ and $h_{1}(1900)$, respectively.
Note that we write $M_{\rm inv}(\phi\pi^0)$ to keep the same nomenclature as in Ref.~\cite{BESIII:2023zwx}, but it is actually $M_{\rm inv}(K^+K^-\pi^0)$ with the cut of Eq.~\eqref{eq:BEScut}. In the large tree contribution, the $K^+, K^-$ do not correspond to the $\phi$ since they are in $I=1$. This corresponds to the ``non-$\phi$'' contribution of Ref.~\cite{BESIII:2023zwx}.}
\label{Fig:Fig4} 
\end{figure}

\begin{figure}[t]
\centering
\includegraphics[width=0.49\textwidth]{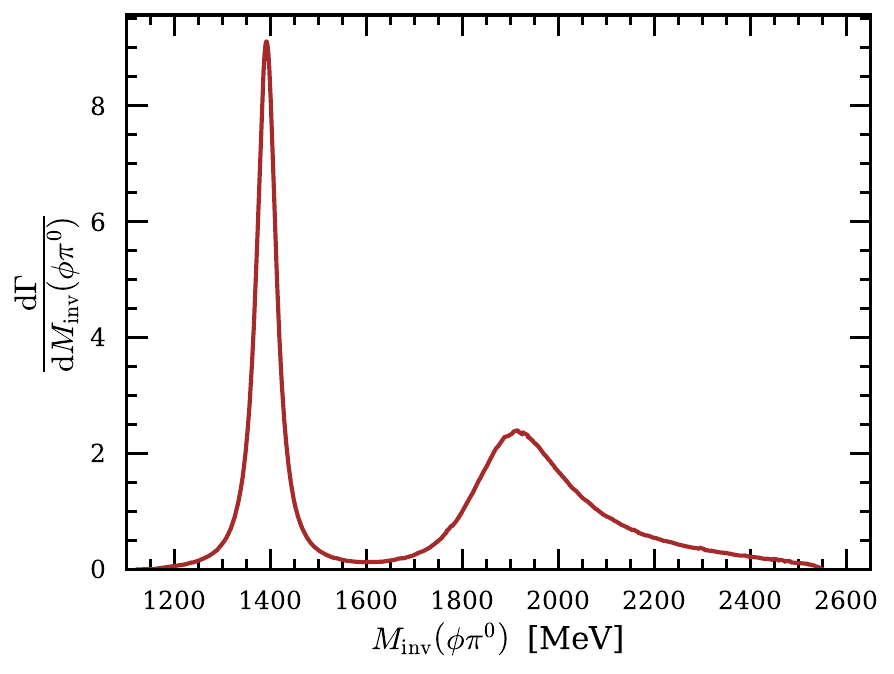}
\vspace{-0.45cm}
\caption{Only triangle singularity result for the $\phi\pi^{0}$ invariant mass distribution.}
\label{Fig:Fig4p}
\end{figure}

In Fig.~\ref{Fig:Fig4} we show the $\phi\pi^{0}$ mass distribution including all the mechanisms discussed in the work.
The data show two clear peaks, one around 1400 MeV and the other around 2100 MeV.
The contribution of the two peaks of the $\phi\eta$ mass distribution (see Fig.~\ref{Fig:Fig6}) only affect the $\phi \pi^0$ mass distribution at high energy.
This allows us to see clearly what is the origin of the two peaks in the $\phi\pi^0$ mass distribution.
Even if the first peak appears close to the predicted position of a triangle singularity coming from the triangle mechanism considered here (Fig.~\ref{Fig:Fig2}), its origin is different.
In fact it comes from the tree level $K^{+}K^{-}$ production of Fig.~\ref {Fig:Fig3}.
The peak is reasonably reproduced using a parameter $A_{tT,890}=3.12\times 10^4$ to adjust to the counts in the experiment.
We also obtain reasonably the second peak by using the same tree level mechanism of Fig.~\ref{Fig:Fig3} and the $K^{*}(1410)$ resonance, using $A_{t,1410}=6.25\times10^4$ to adjust to the counts in the experiment.
This exercise clarifies the origin of these two peaks, that in the experimental analysis of Ref.~\cite{BESIII:2023zwx} are attributed to the ``non-$\phi$" contribution without further detail.
Given the fact that the triangle mechanism of Fig.~\ref{Fig:Fig2} uses the same input as the tree level with the $K^{*}(890)$ resonance, we can now evaluate its contribution to the $\phi\pi^{0}$ mass distribution in the same scale as Fig.~\ref{Fig:Fig4}.
The results are shown in Fig.~\ref{Fig:Fig4p}, where we see that, indeed, there is a peak at $1390 \mev$ corresponding to the predicted triangle singularity, but its strength is about $40$ times smaller than the strength of the experimental peak appearing at the same energy.
There is a second peak associated with the triangle diagram of Fig.~\ref{Fig:Fig2}, where we consider the intermediate $K^{*}$ as the $K^{*}(1410)$.
Its strength is smaller than the one around 1400 MeV and does not show up in Fig.~\ref{Fig:Fig4}.
We can see that there is indeed a triangle singularity associated with the second peak of the mass distribution, appearing at the same position as the experimental peak but it is wider and with smaller strength.

\begin{figure}[t]
\centering
\includegraphics[width=0.49\textwidth]{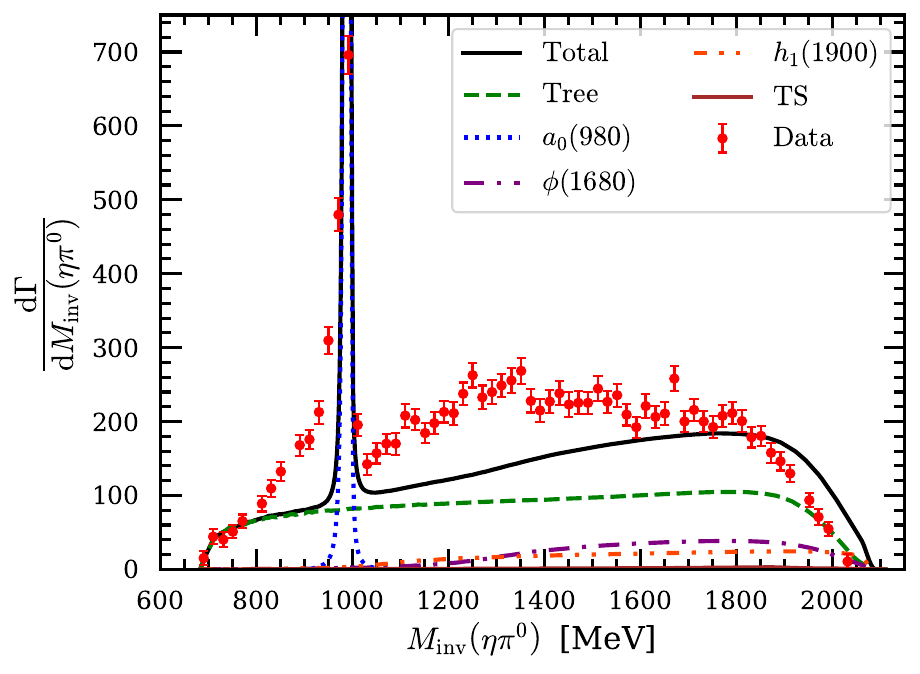}
\vspace{-0.45cm}
\caption{Results for the $\eta\pi^{0}$ invariant mass distribution. The line styles and corresponding contributions are the same as those in Fig.~\ref{Fig:Fig4}.}
\label{Fig:Fig5}
\end{figure}

In Fig.~\ref{Fig:Fig5} we show the $\pi^{0}\eta$ mass distribution.
We see indeed a very sharp peak around the $K\bar{K}$ threshold corresponding to the $a_{0}(980)$.
This contribution comes from the term $t_{a_{0}}$, corresponding to Eq.~\eqref{eq:ta0} and Fig.~\ref{Fig:Fig1}.
Its strength determines the $A_{a_{0}}$ parameter that gets the value, $A_{a_{0}}=1.4\times10^4$.
It might be surprising to see that the signal of the $a_{0}$ is so narrow, both experimentally and theoretically, compared with typical signals observed in isospin allowed channels.
The reason is that the width is actually a measure of the mass difference between the charged and neutral mesons and not a measure of the $a_{0}$ natural width~\cite{Achasov:2002hg,Wu:2007jh,Hanhart:2007bd,Aceti:2012dj,Roca:2012cv}.
We also can see that the largest strength in the $\eta\pi^{0}$ distribution comes from the other mechanisms, in particular the tree level contributions, but they provide a small and smooth background below the peak of the $a_{0}(980)$.

\begin{figure}[t]
\centering
\includegraphics[width=0.48\textwidth]{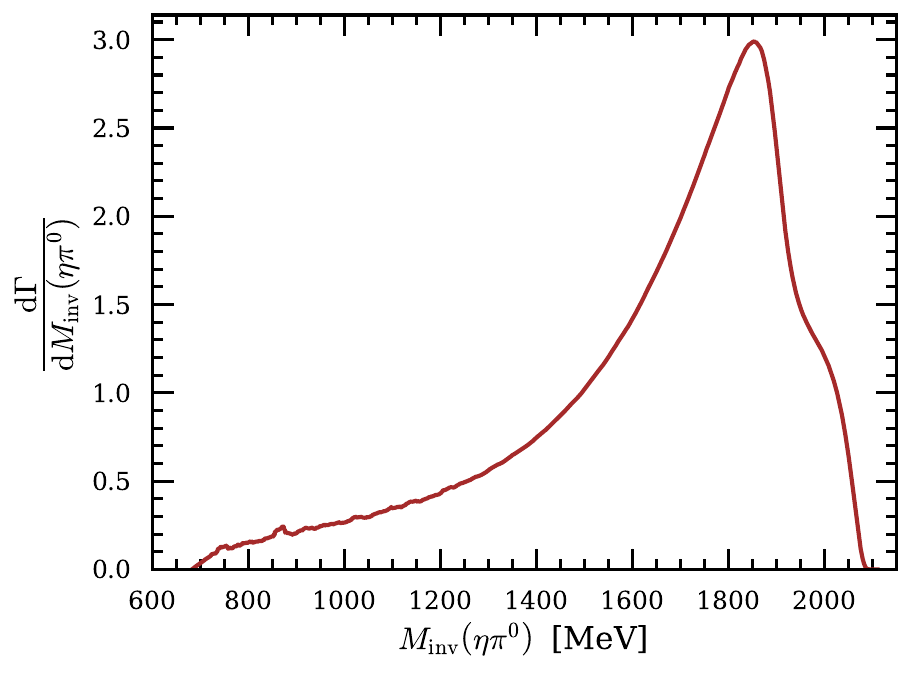}
\vspace{-0.45cm}
\caption{Only triangle singularity result for the $\eta\pi^{0}$ invariant mass distribution.}
\label{Fig:TS-etapi}
\end{figure}

\begin{figure}[t]
\centering
\includegraphics[width=0.49\textwidth]{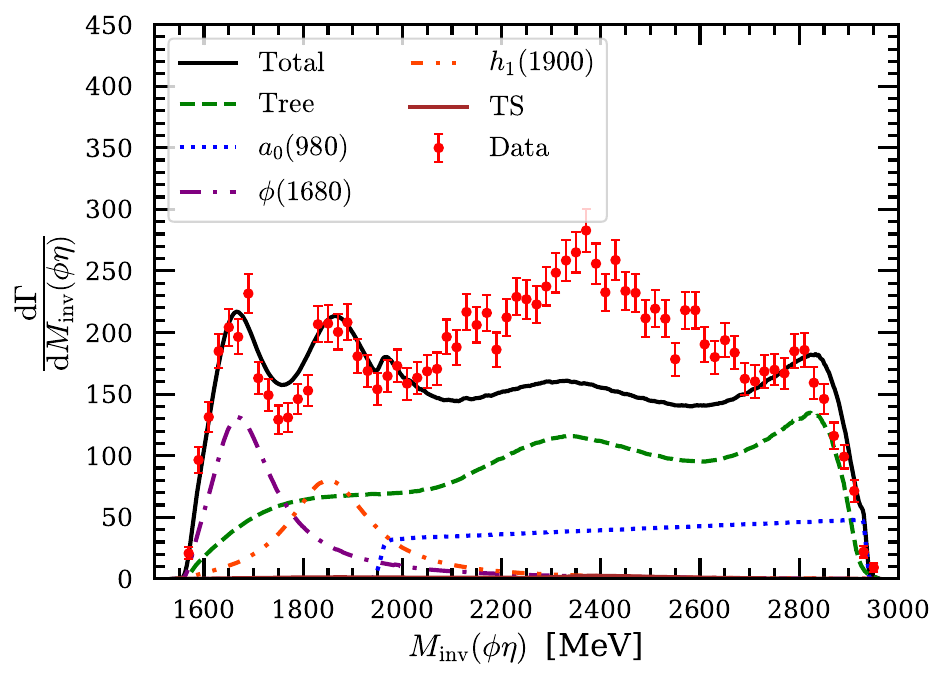}
\vspace{-0.45cm}
\caption{Results for the $\phi\eta$ invariant mass distribution. The line styles and corresponding contributions are the same as those in Fig.~\ref{Fig:Fig4}.}
\label{Fig:Fig6}
\end{figure}

In Fig.~\ref{Fig:Fig5}, the
TS curve is basically invisible. To show its line shape, we plot it in Fig.~\ref{Fig:TS-etapi}. 

We miss some strength in the $\eta \pi^0$ mass distributions already observed in Fig.~\ref{Fig:Fig4}, but this is better seen in the $\phi\eta$ mass distribution that we discuss below.

\begin{figure}[t]
\centering
\includegraphics[width=0.48\textwidth]{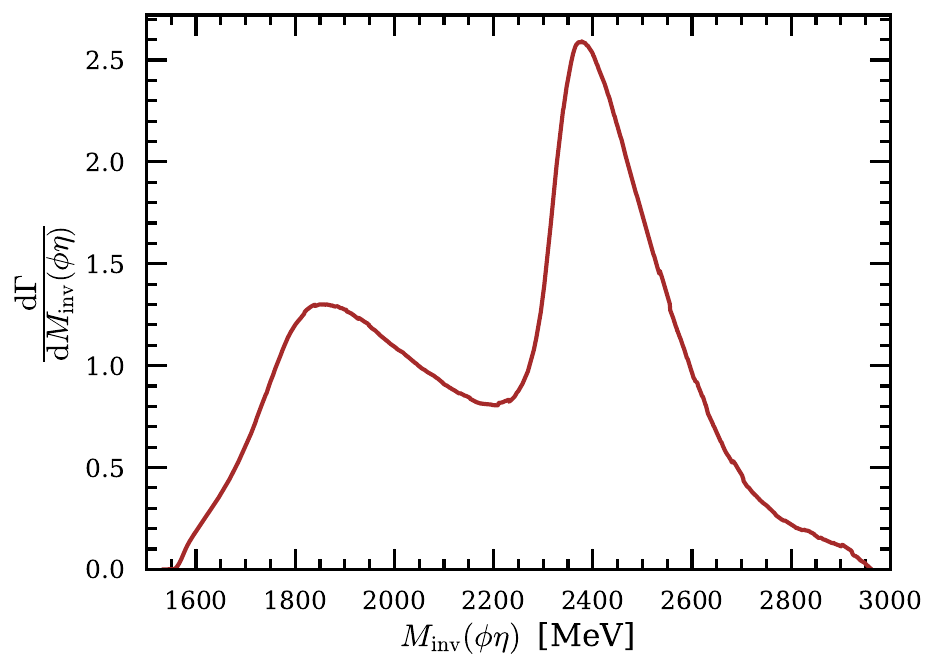}
\vspace{-0.45cm}
\caption{Only triangle singularity result for the $\phi\eta$ invariant mass distribution.}
\label{Fig:TS-phieta}
\end{figure}

We show the $\phi\eta$ mass distribution in Fig.~\ref{Fig:Fig6} and only the TS contribution in Fig.~\ref{Fig:TS-phieta}.
As discussed above, we have fitted the two peaks at around $1650$ and $1850 \mev$ with Breit-Wigner distributions, so we must pay attention to the other structure appearing around $2400 \mev$, as a broad peak, that, in principle, should come from the mechanisms that we have used to describe the data.
We see indeed, that a large fraction of the strength of this peak comes from the tree level contributions, which also produce a small peak around 2800 MeV, as seen in the experiment.
We also see that the tree level contribution produces a broad bump around 2400 MeV where the experimental peak appears.
However, this bump is not sufficient to reproduce the strength of the experimental peak.
It is not clear what the reason for this discrepancy can be, or whether it is the signal of some resonance.
However, the non-$\phi$ contribution of the experimental analysis of Ref.~\cite{BESIII:2023zwx}, reproduces the peak, and we see that the missing strength in this region is reflected in the $\phi\pi^{0}$ mass distribution of Fig.~\ref{Fig:Fig4} in missing strength around $1700 \mev$ between the two peaks coming from the tree level contributions with intermediate $K^*(890)$ and $K^*(1410)$ resonances.
Hence, most likely it reflects the shape of $K^*(890)$ and $K^*(1410)$ resonances obtained assuming propagators with a constant width.
Whatever the reason, we tend to think that this is not a signal of a new state.
It could also be affected by effects on the $h_1(1415)$ resonance [before called $h_1(1380)$] from reinteraction of $K^*\bar K$ in Fig.~\ref{Fig:Fig3} (see Ref.~\cite{Roca:2005nm}).
These effects at the peak of the $\phi\pi^0$ mass distribution around $1400\mev$ are effectively incorporated in our $A_{tT, 890}$ constant and are the same in the triangle diagram of Fig.~\ref{Fig:Fig2} at the peak of the triangle singularity.
Hence the ratio of the strength of these two peaks is well calculated, but there would be corrections in the tree level contribution at higher invariant mass that would modify the shape at Fig.~\ref{Fig:Fig4} in the dip region.

At this point it is interesting to discuss why the peak of the TS and the non-$\phi$ distributions associated with the $K^*(890)$ tree level contribution appear at the same energy in the $M_{\phi\pi^0}$ distribution.
Indeed, as we discussed, a TS in the diagrams of Fig.~\ref{Fig:Fig2} appears when all the three intermediate particles are on shell and are collinear.
This situation will be repeated in the diagrams of Fig.~\ref{Fig:Fig3}, since the $K^+$ and $K^-$ are external particles and hence, on shell.
And in this case the $K^*$ propagator will be placed on shell, and have its maximum, even being infinite if the width is neglected.
The argument can also be repeated for the contribution of the $K^*(1410)$.
In this case, related to the peak observed from the tree level $K^+K^-$ production term, one also finds a TS from the diagrams of Fig.~\ref{Fig:Fig2}, with the substitution $K^* \to K^*(1410)$, peaking at the same place as the tree level but with a much smaller strength.

It has become clear that the non-$\phi$ contribution in this reaction blurs the signals of the TS.
This is due to the experimental mode used in Ref.~\cite{BESIII:2023zwx} to identify the $\phi$, by looking at its $K^+K^-$ decay.
If one identified the $\phi$ with another method or other decay mode, say three pions for instance, this undesired background would disappear, allowing one to observe the TSs that are unavoidable, since their strength is tied to the non-$\phi$ peak observed in Ref.~\cite{BESIII:2023zwx} that we have fitted, and thus our predictions are reliable.
A different way to observe the TS related to this process is by looking at the $J/\psi \to \phi \pi^+ a_0(980)$ reaction, discussed in Ref.~\cite{Xiao:2024ohf}.

The present work has served to show that, indeed, the isospin forbidden mode $J/\psi \to \phi \pi^0 \eta$ exists and leads to a clean signal for the $a_0(980)$ resonance, and at the same time has clarified the origin of other peaks seen in the $\phi\eta$ and $\phi \pi^0$ mass distributions, which should not be attributed to new resonances nor to TS contributions.

\section{FURTHER INTERACTION STEPS}
We have $J/\psi$ decay to four particles $K^+K^- \eta \pi^0$ and we have considered interactions of pairs.
In Fig.~\ref{Fig:Fig1} we have $K^+ K^-$ coming from the $\phi$ decay, and $K\bar K \to \pi^0 \eta$.
In Fig.~\ref{Fig:Fig3} we have $K^*\to \pi K$.
In the triangle diagrams of Fig.~\ref{Fig:Fig2} we have a three-body interaction, $K^* \to \pi K$ and the $K$ coming from this decay interacts with a third particle.
We found that this mechanism, even producing a triangle singularity, gives a very small signal.

In the case of three particles in the final state, there is the traditional approach of Khuri-Treiman \cite{Khuri:1960zz,Guo:2014vya} which accounts for multiple scattering of particles, accounting for the interaction of a pair followed by reinteraction of one of these particles with the third particle.
Our triangle diagrams of Fig.~\ref{Fig:Fig2} would be one example of interaction beyond the interaction of just pairs.
The Khuri-Treiman approach is of much value when studying the decay of light particles, where there is little phase space and the kinematical conditions in rescattering are not very different from those of the first scattering.
Such is the case in $\eta \to 3 \pi$ \cite{Guo:2015zqa}, the $\omega/\phi \to 3 \pi$ \cite{Danilkin:2014cra} or the $\phi \to 3 \pi$ in Ref.~\cite{Garcia-Lorenzo:2025uzc}.
In decays of heavy hadrons, there is more phase space and the scattering of a pair normally selects some resonance. 
Fixing the two-body kinematics to this resonance, the range of invariant mass of one of the particles of the pair with the third particle has limitations that make it unlikely to match with another resonance for the new pair, the only way to make competitive the new mechanism with an extra loop.
This has been discussed in detail in Sec. IV A of Ref.~\cite{Song:2025dhb} in the $\Lambda_b \to D^+ D^- \Lambda$, $D^0 D_s^- p$, $D_s^+ D_s^- \Lambda$ reactions.
Yet, we find it instructive to discuss the consequence of going one step further in the interaction in the mechanisms that we have.

\begin{figure}[t]
\centering
\includegraphics[width=0.47\textwidth]{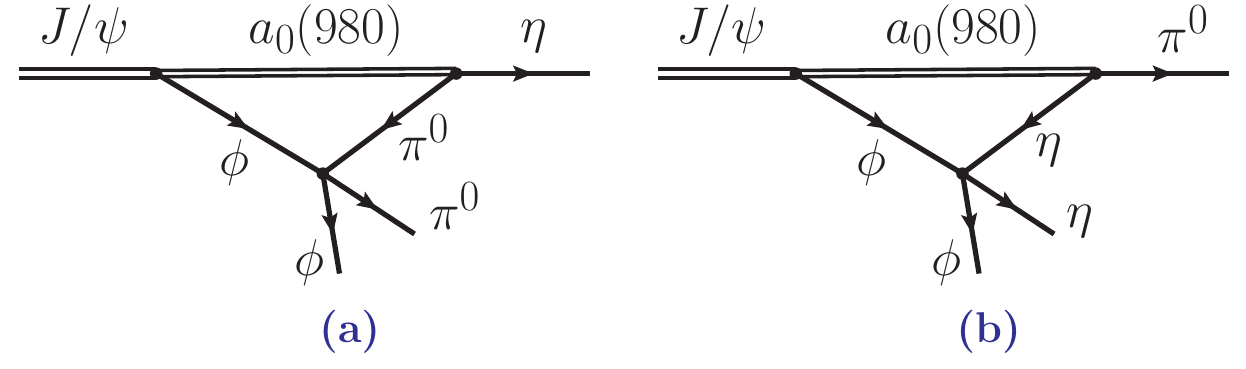}
\vspace{-0.45cm}
\caption{Rescattering mechanisms from Fig.~\ref{Fig:Fig1}: (a) $\phi \pi^0$ interaction generating the $b_1(1235)$ resonance; (b) $\phi \eta$ interaction generating the $h_1(1170)$, $h_1(1415)$ resonances.}
\label{Fig:new1}
\end{figure}

Let us start from the mechanism of Fig.~\ref{Fig:Fig1}, where the $K\bar K \to \pi^0 \eta$ interaction proceeds through the excitation of the $a_0(980)$ resonance, as shown in Fig.~\ref{Fig:Fig5}.
We can depict the rescattering mechanisms as shown in Fig.~\ref{Fig:new1}.
These are triangle diagrams, which could develop or not a triangle singularity.
From the experience of the results from the triangle diagrams of Fig.~\ref{Fig:Fig2}, we might expect a small contribution with respect to the large ones evaluated before, even more if the diagrams of Fig.~\ref{Fig:new1} do not develop a TS.  

To estimate the relevance of the diagrams of Fig.~\ref{Fig:new1}, we use two sources of information:
\begin{itemize}
	\item[1)] Which resonances couple to $\phi \pi^0$ and $\phi \eta$? We find this information in the work of Ref.~\cite{Roca:2005nm}.
	\item[2)] How close or far away are we from having a TS in these diagrams? We find this information in Eq.~(18) of Ref.~\cite{Bayar:2016ftu}, which is the test that all particles in the loop are on shell and collinear, satisfying the Coleman-Norton theorem \cite{Coleman:1965xm} that the mechanism can proceed at the classical level.
\end{itemize}
In Ref.~\cite{Roca:2005nm} we find that the $\phi \pi^0$ channel couples strongly to the $b_1(1235)$ resonance.
Direct application of Eq.~(18) of Ref.~\cite{Bayar:2016ftu} to the diagram of Fig.~\ref{Fig:new1}(a) tells us that a TS develops if the mass of the $J/\psi$ was $2027 \mev$, very far away from the actual mass of the $J/\psi$, which makes that mechanism negligible.

Concerning the diagram of Fig.~\ref{Fig:new1}(b), the $\phi \eta$ channel couples to the $h_1$ resonances, $h_1(1170)$ and $h_1(1380)$ [now $h_1(1415)$].
However, as found in Ref.~\cite{Roca:2005nm}, the coupling of $\phi \eta$ to the $h_1(1170)$ is extremely small, and hence only the $h_1(1415)$ that has an appreciable coupling to $\phi \eta$ could be relevant.
Once again, application of Eq.~(18) of Ref.~\cite{Bayar:2016ftu} tells us that a TS develops if the mass of the $J/\psi$ was $2049\mev$, far away from the actual mass, making again the contribution of this mechanism negligible.

\begin{figure}[t]
\centering
\includegraphics[width=0.27\textwidth]{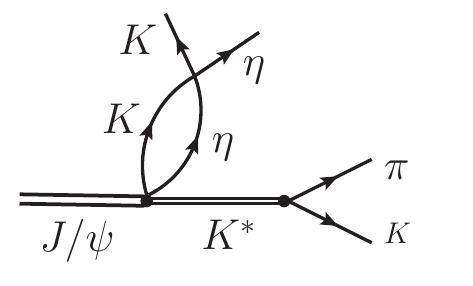}
\vspace{-0.45cm}
\caption{Rescattering mechanisms of Fig.~\ref{Fig:Fig3} through $\eta K$ interaction.}
\label{Fig:new2}
\end{figure}

Next we turn our attention to the important diagrams of Fig.~\ref{Fig:Fig3} and proceed with possible rescatterings.
The $K\bar K$ rescattering leads to the triangle mechanisms of Fig.~\ref{Fig:Fig2}, which develop a TS, in spite of which, the contribution is very small.
Then we consider the $K\eta$ interaction.
This leads us to the mechanism depicted in Fig.~\ref{Fig:new2}.
This would add a term with the former amplitude multiplied by $G_{K\eta}\; t_{K\eta,\, K\eta}$.
The $K\eta$ couples to the $K_0^*(700)$ resonance, but its coupling is smaller than to $\pi K$, and appears squared in $t_{K\eta,\, K\eta}$, rendering a factor of about 3 reduction with respect to $t_{\pi K,\, \pi K}$ \cite{Oller:1998zr}, but more important, $m_{K}+m_\eta=1041\mev$, far away from the nominal mass of the $K_0^*(700)$ even considering its large width.
Hence, this contribution is also very small.

\begin{figure}[t]
\centering
\includegraphics[width=0.45\textwidth]{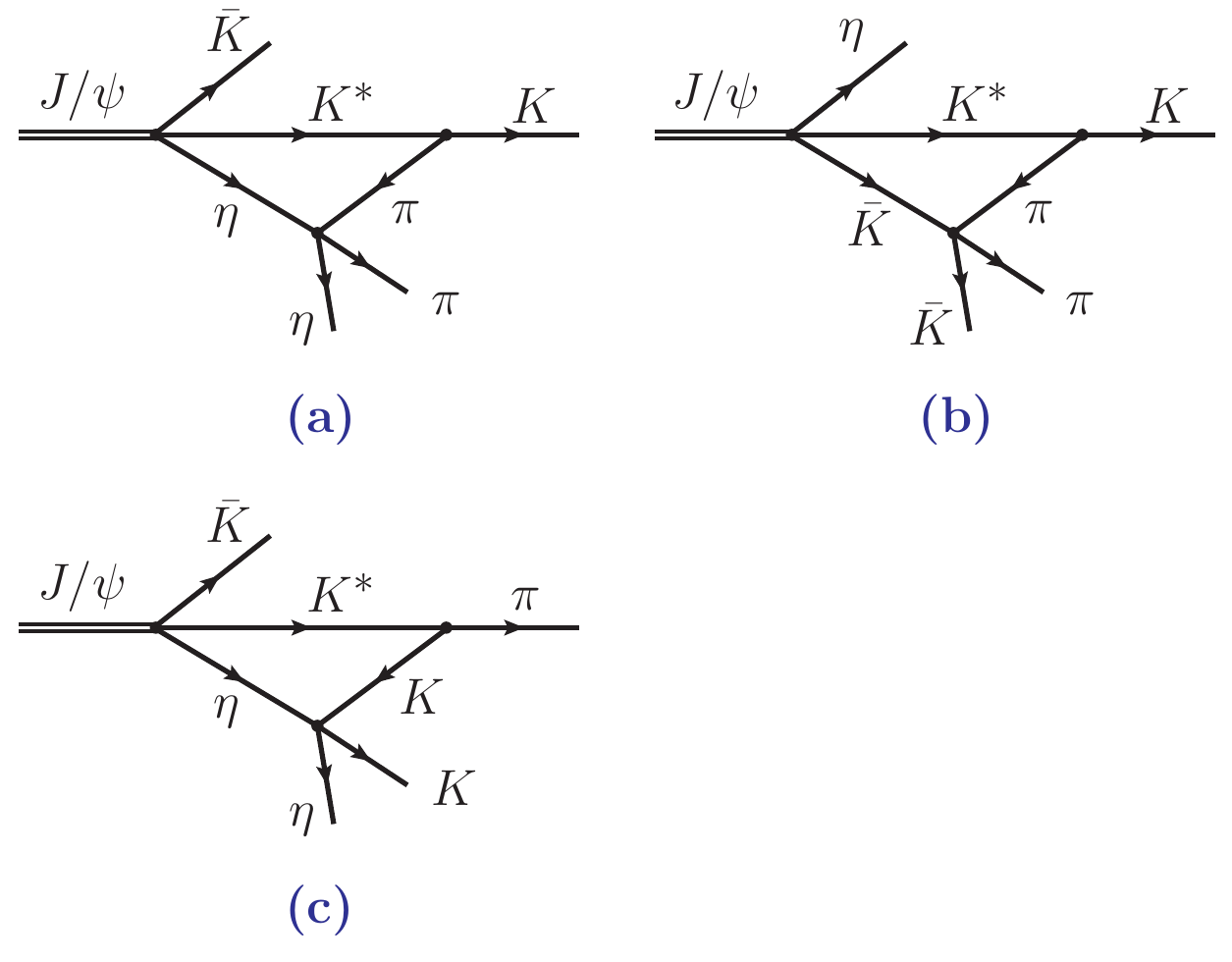}
\vspace{-0.45cm}
\caption{Rescattering mechanisms from Fig.~\ref{Fig:Fig3} leading to triangle diagrams: (a) with $\pi \eta$ rescattering; (b) with $\pi \bar K$ rescattering; (c) with $\eta K$ rescattering.}
\label{Fig:new3}
\end{figure}

Next we make the $K^+$ or the $\pi^0$ of Fig.~\ref{Fig:Fig3}(a) reinteract with the $\eta$ or $K^-$ and we get three new mechanisms depicted in Fig.~\ref{Fig:new3}.

The diagram of Fig.~\ref{Fig:new3}(a) would produce the $\eta \pi$ with the $a_0(980)$ resonance.
Once again, applying Eq.~(18) of Ref.~\cite{Bayar:2016ftu}, we find that one is extremely far away from satisfying the conditions of a TS and, again, the mechanism is totally negligible.
In Fig.~\ref{Fig:new3}(c), the $\eta K$ couples to the $K^*_0(700)$ but it cannot develop a TS because the $\eta K$ mass exceeds the one of the $K^*_0(700)$ by far, even considering the width; in addition, as already discussed above, the amplitude $\eta K \to \eta K$ through the resonance is very small for the actual energies of $\eta$ and $K$ in the phase space.

Finally, we turn to the mechanism of Fig.~\ref{Fig:new3}(b).
This case requires a bit more attention. The $\bar K \pi$ can now couple to the $K^*_0(700)$ and there is no problem placing the $\bar K \pi$ on shell in the loop.
Then the $\bar K \pi$ pair can sit on top of the $K^*_0(700)$ resonance.
Then we apply Eq.~(18) of Ref.~\cite{Bayar:2016ftu} and find that only in the range of $640$-$760 \mev$ for $M_{\rm inv}(\bar K\pi)$ we can find a TS.
This restricts the mass range of the $K^*_0(700)$ to about $1/3$ given that $\Gamma_{K^*_0(700)} \sim 500 \mev$.
However, more interesting, for $M_{\rm inv}(\bar K\pi)=640 \mev$ the TS appears at 
$M_{\rm inv}(K \bar K\pi)=1584 \mev$ \footnote{$M_{\rm inv}(K \bar K\pi)$ would correspond to $M_{\rm inv}(\phi\pi)$ in Fig.~\ref{Fig:Fig4}.}.
Similarly, if we take $M_{\rm inv}(\bar K\pi)=760\mev$, the TS appears at $M_{\rm inv}(K \bar K\pi)=1383 \mev$.
Thus, we see that, in the limited range of $M_{\rm inv}(\bar K\pi)$ where a TS can appear, the equivalent $M_{\rm inv}(\phi \pi)$ where the TS appears moves within $200\mev$.
This indicates that this TS would have a width of this size, much bigger than the TS from the diagram of Fig.~\ref{Fig:Fig2}, where the small width of the $\phi$ made the peak of the TS much narrower.
We could expect a broad and small peak around the same region as the one obtained from the mechanism of Fig.~\ref{Fig:Fig2}, with a smaller strength and split in a broader region which would still allow one to identify the peak of Fig.~\ref{Fig:Fig4p}.
Note also that the mechanism of Fig.~\ref{Fig:new3}(b) does not correspond to $\phi \pi$; hence, it would also not contribute in a reaction where one measures the genuine $\phi \pi \eta$, the channel suggested to observe the TS of Fig.~\ref{Fig:Fig4p}.

\section{Conclusions}
We have carried out an analysis of the $J/\psi \to \phi \eta \pi^0$ reaction recently measured with high precision by the BESIII Collaboration \cite{BESIII:2023zwx}, where the $a_0(980)$ resonance, which appears in an isospin forbidden mode, is clearly seen.
Explaining this production mode is one of the purposes of the present work.
Another motivation was the prediction in Ref.~\cite{Jing:2019cbw} that there should be a triangle singularity from a reaction mechanism, showing a peak in the $\phi\pi^0$ mass distribution around $1385\mev$~\cite{Jing:2019cbw}.
The experimental data show a clear peak around this energy, which, however, was identified in the experimental analysis as a ``non-$\phi$'' contribution.
Trying to decipher the origin of this ``non-$\phi$'' contribution is also an issue that we tackle in the present work.

We described three mechanisms contributing to the reaction.
The first one consists in producing $J/\psi \to \phi K^+K^-, \phi K^0 \bar K^0$ allowing the $K^+K^-$ and $K^0 \bar K^0$ to undergo final state interaction to get $\pi^0 \eta$ and hence a signal of the $a_0(980)$.
This mechanism would not produce $\pi^0 \eta$ at the end if the masses of the charged and neutral kaons were equal, but as soon as physical masses are used in the loops, isospin symmetry is violated, and the $a_0(980)$ production is possible.
Yet, since this reaction is made possible by the mass difference between $K^+$ and $K^0$, the shape of the $\pi^0 \eta$ mass distribution around the $K\bar K$ threshold is very narrow, reflecting this mass difference rather than the natural width of the $a_0(980)$.
This feature is supported by the data and follows predictions from previous theoretical studies.

We also found the origin of the peaks in the $\phi\pi^0$ mass distribution as coming from the way the $\phi$ was identified in the experiment measuring $K^+K^-$ in a narrow window around the $\phi$ mass.
However, we could show that there was a tree level mechanism for $J/\psi \to \eta K^+ K^- \pi^0$, where the pair $K^+, K^-$ was produced in isospin $I=1$, hence not coming from $\phi$ decay, and provided a contribution to the mass distribution.
The mechanism for this tree level contribution is $J/\psi \to \eta K^- K^{*+}; \,K^{*+} \to \pi^0 K^+$ plus $J/\psi \to \eta K^+ K^{*-}; \,K^{*-} \to \pi^0 K^-$.
If the intermediate $K^*$ is $K^*(890)$, a peak appears around $1390\mev$, and if the intermediate $K^*$ is $K^*(1410)$, a peak appears around $2050\mev$, and both peaks are seen in the experimental data.

Related to these latter mechanisms there is a triangle mechanism, suggested in Ref.~\cite{Jing:2019cbw}, where the $K^+$ and $K^-$ merge to produce $\pi^0 \eta$ and hence the $a_0(980)$.
The strength of this mechanism is tied to that of the tree level one, since only the $K\bar K \to \pi^0 \eta$ transition amplitudes, well known theoretically, are needed to complete the loop diagram.
Then we can make a good prediction of the strength of this mode, also violating isospin, which turns out to be about a factor $40$ smaller than the peaks that originated from the tree level mechanisms that do not involve isospin violation.

We also indicate that the triangle singularities could possibly be observed in the same reaction, but identifying the $\phi$ through different decay channels than the $K^+K^-$ chosen in the BESIII experiment.

In addition, we made a thorough discussion of possible effects of extra interactions beyond two-body  interactions and found the contributions negligible.

In summary, we provided an explanation for the $a_0(980)$ production mode, with the abnormally narrow width observed in the experiment.
We also could identify the origin of the peaks in the $\phi\pi^0$ mass distribution observed in the experiment, branded as ``non-$\phi$'' contribution in the experimental analysis; finally we could see that there are indeed two peaks in the $\phi\pi^0$ mass distribution showing a triangle singularity, but their strength is significantly smaller than that of the observed peaks.
We also mentioned that to see these peaks the technique used so far to identify the $\phi$, by looking at its $K^+K^-$ decay, would have to be changed to avoid contributions from $K^+K^-$ that do not come from $\phi$ decay and have a large strength. The results obtained here should be of use for planning future experiments aiming at identifying the triangle singularities predicted here.

\section*{Acknowledgments}
We would like to acknowledge useful discussions with F. K. Guo, X. R. Lyu, B. S. Zou, and C. Z. Yuan.
This work is partly supported by the National Natural Science Foundation of China (NSFC) under Grants No. 12575081 and No. 12365019,
and by the Natural Science Foundation of Guangxi province under Grant No. 2023JJA110076,
and by the Central Government Guidance Funds for Local Scientific and Technological Development, China (No. Guike ZY22096024),
and partly by the Natural Science Foundation of Changsha under Grant No. kq2208257 and the Natural Science Foundation of Hunan province under Grant No. 2023JJ30647.
This work is also partly supported by the Spanish Ministerio de Economia y Competitividad (MINECO) and European FEDER funds under Contracts No. FIS2017-84038-C2-1-PB, PID2020-112777GB-I00, and by Generalitat Valenciana under contract PROMETEO/2020/023. 
This project has received funding from the European Union Horizon 2020 research and innovation program under the program H2020-INFRAIA-2018-1, grant agreement No. 824093 of the STRONG-2020 project.

\appendix
\section{Four-body phase space calculation and invariant mass distributions}

We follow the step of Ref.~\cite{Song:2022kac}.
The decay width for $J/\psi \to\pi^0 \eta K^+K^-$ is given by
\begin{align}\label{eq:Appen1}
	\Gamma
	=&\dfrac{1}{2\, M_{J/\psi}} \int \dfrac{d^3p_\eta}{(2\pi)^3} \; \dfrac{1}{2\, E_\eta} \int \dfrac{d^3p_{K^+}}{(2\pi)^3} \; \dfrac{1}{2\, E_{K^+}}
	\nonumber \\
	&\times \int \dfrac{d^3p_{K^-}}{(2\pi)^3} \; \dfrac{1}{2\, E_{K^-}} \int \dfrac{d^3p_{\pi^0}}{(2\pi)^3} \; \dfrac{1}{2\, E_{\pi^0}}
	\nonumber \\
	&\cdot (2\pi)^4 \;\delta^4(P-p_\eta-p_{K^+}-p_{K^-}-p_{\pi^0}) \;|t|^2,
\end{align}
with $P$ the momentum of the $J/\psi$, which we take at rest. The $d^3p_{\pi^0}$ integral is killed with a $\delta^3(\vec P-\vec p_\eta-\vec p_{K^+}-\vec p_{K^-}-\vec p_{\pi^0})$, which gives
\begin{equation}
	\vec p_{\pi^0}=-(\vec p_\eta+\vec p_{K^+}+\vec p_{K^-}).
\end{equation}
We introduce the variables
\begin{equation}
	\begin{cases}
		\vec{P}_{K}=\vec{p}_{K^{+}}+\vec{p}_{K^{-}},\\[1mm]
		\vec{q}_{K}=\vec{p}_{K^{+}}-\vec{p}_{K^{-}},\\
	\end{cases}
	\text{or}~~
	\begin{cases}
		\vec{p}_{K^{+}}=\frac{1}{2}(\vec{P}_{K}+\vec{q}_{K}),\\[1mm]
		\vec{p}_{K^{-}}=\frac{1}{2}(\vec{P}_{K}-\vec{q}_{K}).
	\end{cases}
\end{equation}
Then we have
\begin{align}\label{eq:Appen4}
	\Gamma
	=&\dfrac{1}{2\, M_{J/\psi}}\,\frac{1}{8} \int \dfrac{d^3p_\eta}{(2\pi)^3} \; \dfrac{1}{2\, E_\eta} \int \dfrac{d^3P_{K}}{(2\pi)^3} \, \int \dfrac{d^3q_{K}}{(2\pi)^3}
	\nonumber \\
	&\cdot  \frac{1}{2\, E_{K^{+}}}\,\dfrac{1}{2\, E_{K^-}}\,\dfrac{1}{2\, E_{\pi^0}}
	\nonumber \\
	&\cdot 2\pi \; \delta(M_{J/\psi}-E_{\eta}-E_{K^{+}}-E_{K^{-}}-E_{\pi^{0}}) \;|t|^2.
\end{align}
Then
\begin{equation}
	\vec{p}_{\pi^{0}}=-(\vec{p}_{\eta}+\vec{P}_{K}).
\end{equation}

We take advantage of the rotational invariance of the final system and take
\begin{equation}\label{eq:Appen6}
	\vec{p}_{\eta}=p_{\eta}
	\begin{pmatrix}
		0\\0\\1
	\end{pmatrix},\quad
	\vec{P}_{K}=P_{K}
	\begin{pmatrix}
		\sin\theta\cos\phi\\
		\sin\theta\sin\phi\\
		\cos\theta
	\end{pmatrix}.
\end{equation}
Then we eliminate the $\delta(M_{J/\psi}-E_{\eta}-E_{K^{+}}-E_{K^{-}}-E_{\pi^{0}})$ of energy by integrating over $\cos\theta$, which fixes it to be
\begin{align}
	\nonumber
	A_{\theta}=\cos\theta=\frac{1}{2P_{K}\;p_{\eta}}
	\Big\{&(M_{J/\psi}-E_{\eta}-E_{K^{+}}-E_{K^{-}})^2\\
		   &-m_{\pi^{0}}^2-\vec{p}\,^2_{\eta}-\vec{P}^2_K\Big\},
\end{align}
but the phase space requires
\begin{equation}
	\Theta(1-A^2_{\theta})\;\Theta(M_{J/\psi}-E_{\eta}-E_{K^{+}}-E_{K^{-}}).
\end{equation}
We define $\tilde{q}_K$ referred to the $\vec{P}_{K}$ direction, as if $\vec{P}_{K}$ were in the $z$ direction,
\begin{equation}
	\tilde{q}_{K}=q_{K}
	\begin{pmatrix}
		\sin\tilde{\theta}_{q}\cos\tilde{\varphi}_{q}\\
		\sin\tilde{\theta}_{q}\sin\tilde{\varphi}_{q}\\
		\cos\tilde{\varphi}_{q}
	\end{pmatrix},
\end{equation}
and perform the rotation that leads $\vec{P}_{K}$ from the $z$ direction to the actual $\vec{P}_{K}$ in the original $
J/\psi$ rest frame where Eqs.~\eqref{eq:Appen6} are defined.
We obtain $\vec{q}_{K}$ in the original frame by means of the rotation
\begin{equation}
	\vec{q}_{K}=R\;\tilde{q}_{K},
\end{equation}
where the rotation matrix $R$ is given by
\begin{equation}
	R=\begin{pmatrix}
		\cos\varphi\cos\theta&-\sin\varphi&\cos\varphi\sin\theta\\
		\sin\varphi\cos\theta&\cos\varphi&\sin\varphi\sin\theta\\
		-\sin\theta&0&\cos\theta
	\end{pmatrix}.
\end{equation}
Then the final formula for $\Gamma$ is given by
\begin{align}
	\nonumber
	\Gamma=&\dfrac{1}{M_{J/\psi}}\;\dfrac{1}{64\,\pi}\dfrac{1}{(2\pi)^6}\int \dd E_{\eta}\int P_{K}\;\dd P_{K}\,\dd\varphi\\
	\nonumber
	       &\cdot\int \tilde{q}^2_{K}\, \dd\tilde{q}_{K}\;\dd\cos\tilde{\theta}_{q} \; \dd\tilde{\varphi}_{q} \;\dfrac{1}{2\,E_{K^{+}}}\;\dfrac{1}{2\,E_{K^{-}}}\\
	       &\cdot \abs{\tilde{t}}^2\;\Theta(1-A^2_{\theta})\;\Theta(M_{J/\psi}-E_{\eta}-E_{K^{+}}-E_{K^{-}}).
\end{align}
The rest of the work is done using standard Monte Carlo (MC) calculations generating events over an enlarged phase space that contains the actual phase space and impose the $K^{+}K^{-}$ cut of Eq.~\eqref{eq:BEScut}.
For every good MC event, different invariant masses are evaluated and stored in boxes, which at the end provide the differential mass distributions.
Apart from the trivial limits of the different variables, it is worth mentioning the maximum values of $P_{K}$ and $q_{K}$.
The former corresponds to the case when the $K^{+}K^{-}$ and $\pi^{0}\eta$ pairs have their minimum invariant masses.
Thus,
\begin{equation}
	P_{K,\text{max}}=\frac{\lambda^{1/2}(M_{J/\psi}^2,(m_{K^{+}}+m_{K^{-}})^2,(m_{\pi^{0}}+m_{\eta})^2)}{2M_{J/\psi}}.
\end{equation}
The maximum value of $q_{K}$ is obtained when $\pi^{0}\eta$ are produced at rest and the $K^{+}K^{-}$go opposite to each other, in which case
\begin{equation}
	p_{K^{+}}=
	\frac{\lambda^{1/2}((M_{J/\psi}-m_{\pi^0}-m_{\eta})^2,m_{K^{+}}^2,m_{K^{-}}^2)}{2\,(M_{J/\psi}-m_{\pi^0}-m_{\eta})},
\end{equation}
and
\begin{equation}
	\tilde{q}_K=q_{K,\text{max}}=2\,p_{K^{+}}.
\end{equation}

\bibliographystyle{a}
\bibliography{refs}
\end{document}